\documentclass[letter,scriptaddress, twocolumn, showkeys]{revtex4}
\usepackage[utf8]{inputenc}
\usepackage{amsfonts}
\usepackage{amsmath}
\usepackage{amsthm}
\usepackage{amssymb}
\usepackage{csvsimple}
\usepackage[hidelinks]{hyperref}
\hypersetup{
    colorlinks=true,
    linkcolor=blue,
    filecolor=magenta,      
    urlcolor=cyan,
    pdftitle={Overleaf Example},
    pdfpagemode=FullScreen,
    }
\usepackage{booktabs}
\usepackage{graphicx}
\graphicspath{ {images/} }
\usepackage{graphicx, animate}
\usepackage{multimedia}
\usepackage{media9}
\graphicspath{ {./images/} }
\usepackage{wrapfig}
\usepackage{color}
\usepackage{chngcntr}
\counterwithout{table}{section}
\usepackage{algorithm}
\usepackage{algpseudocode}
\usepackage{wrapfig}
\usepackage{caption}
\usepackage{subcaption}
\usepackage{cleveref}

\theoremstyle{definition}
\newtheorem{definition}{Definition}

\theoremstyle{remark}

\begin{document}
\title{Influence Prediction in   Collaboration Networks:\\ An Empirical Study on arXiv}
\author{Marina Lin$^a$, Laura P. Schaposnik$^b$ and Raina Wu$^c$}

\begin{abstract}
This paper provides an empirical study of the Social Sphere Model for influence prediction, previously introduced by the authors, combining link prediction with top‑$k$ centrality-based selection. We apply the model to the temporal arXiv General Relativity and Quantum Cosmology collaboration network, evaluating its performance under varying edge sampling rates and prediction horizons to reflect different levels of initial data completeness and network evolution. Accuracy is assessed using mean squared error in both link prediction and influence maximization tasks. The results show that the model effectively identifies latent influencers - nodes that are not initially central but later influential - and performs best with denser initial graphs. Among the similarity measures tested, the newly introduced RA-2 metric consistently yields the lowest prediction errors. These findings support the practical applicability of the model to predict real-world influence in evolving networks.
\end{abstract}

\maketitle

\section{Introduction}
\label{sec:intro}
Predicting who \emph{will}  influence in the future is a markedly different challenge from identifying who matters today (e.g. see \cite{doi:10.1142/S0219525923500091,listofidentificationtypes} and references therein).  In our previous work \cite{lin2025socialspheremodelheuristic}, we introduced the \textit{\textbf{Social-Sphere Model}}, coupling path-based link-prediction scores with low-cost influence-maximization algorithms, offering an interpretable alternative to deep learning approaches \cite{info13030123, NEURIPS2018_53f0d7c5, hamilton2020grl}.  While the \textit{Social Sphere Model} demonstrated promising results on synthetic networks such as large scale Erdős–Rényi graphs, its real-world applicability remained untested.  Real-world data tends to be noisy, partially-observed, and temporally evolving data as could be citations, friendships, or collaboration networks.

Previous empirical studies on influence maximization have leaned on greedy submodular algorithms~\cite{scalenetworks} or hybrid pipelines that predict future graph structure using link prediction models and apply IM heuristics to the resulting networks~\cite{Yanchenko2023}. Recent advances in link prediction and influence maximization have made significant use of geometric deep learning (GDL), particularly Graph Neural Networks (GNNs). These models learn node embeddings that capture complex structural and temporal features of evolving networks. GNN-based and temporal variants have shown strong performance in tasks such as dynamic link prediction and influence estimation \cite{info13030123, NEURIPS2018_53f0d7c5, hamilton2020grl, rossi2020temporalgraphnetworksdeep}. However, these approaches come at a cost as they typically require extensive training on large datasets and limit interpretability, and they often require temporal labels or timestamps that may not be available. In contrast, the Social Sphere Model relies on interpretable, path-based link prediction scores and top-k heuristics to identify future influencers. Moreover, to the best of our knowledge, no work systematically benchmarks a lightweight heuristic such as SSM across multiple edge-observation regimes to quantify how data sparsity affects predictive accuracy.

In the present paper, we shall consider the particular case of the scientific collaboration graph from the arXiv General Relativity and Quantum Cosmology (GR-QC) coauthorship network containing more than $5{,}000$ researchers and $14{,}000$ weighted edges~\cite{leskovec2008communitystructure} from the Stanford Large Network Dataset Collection. This data set exhibits both heavy-tailed degree distributions and latent ties that form over time, making it a robust choice for modeling real-world situations. Furthermore, failure to account for these latent ties can mislead hiring committees, funding agencies, and outreach campaigns aiming to target emerging leaders rather than established stars. We shall seek to quantify influence through simple and complex contagion \cite{simplecontagion} \cite{complexcontagiondefinition} through our {\it Social Sphere Model} from ~\cite{lin2025socialspheremodelheuristic}. Our study aims to test whether heuristic link-prediction metrics (e.g.\ RA-2) suffice in large-scale datasets and efficiently predict latent influencers. In particular, we shall do the following:

\begin{enumerate}
    \item \textbf{Dataset-driven validation.}  We apply SSM to the GR-QC collaboration network under two observation regimes ($90\%$ and $70\%$ revealed edges) and two forecast horizons ($t\!=\!1$ and $t\!=\!3$), evaluating both link-prediction quality and influence spread under simple and complex contagion models.
    \item \textbf{Latent influencer analysis.}  We introduce an evaluation protocol that distinguishes surface influencers (already highly central) from latent influencers (low initial centrality with high future impact), showing that SSM recovers the latter in up to $75\%$ of trials.
    \item \textbf{Metric benchmarking.}  We benchmark seven similarity metrics; our RA-2 score achieves the lowest mean-squared error (MSE) in $75\%$ of scenarios and the  lowest average MSE across all settings.
\end{enumerate}

Our work is organized in the following way: Section~\ref{sec:background} gives a concise recap of SSM-I  and   Section~\ref{sec:methods} describes our dataset, evaluation pipeline, and description of the models compared.  Results and analyses appear in Section~\ref{sec:results}, followed by discussion and future extensions in Section \ref{sec:conclusion}.

\section{Background}
\label{sec:background}
In this section, we provide the theoretical foundation necessary for our empirical study of influence prediction in temporal networks. We begin by reviewing basic graph-theoretic representations of social networks and key link prediction techniques, including both standard and newly introduced similarity metrics. We then describe models of influence spread, with particular attention to how contagion dynamics and temporal features can affect the role and identification of influential nodes.
\subsection{Social Networks and Link Prediction}
In what follows, we briefly recall the link prediction framework and notational conventions from~\cite{lin2025socialspheremodelheuristic}, which form the basis for the analysis in this paper.
Social networks are modeled using graph theory by expressing the network as a graph $G = (V,E)$, for $V$ the set of vertices (the individuals of the network) and $E$ the set of edges (the connections between individuals).  Given such a graph, one may assign a weight function  
\[
w \colon E \rightarrow [0,1],
\] 
where $w(u,v)$ is the weight of edge $uv$.  In line with Goyal et al.~\cite{bernoullidistribution}, we treat $w(u,v)$ as the empirical probability of influence transmission between $u$ and $v$ during a unit time‐interval. Because a geometric distribution counts the expected number of Bernoulli trials until success, the expected time for an interaction on edge $uv$ is $\tfrac{1}{w(u,v)}$.  We therefore define:

\begin{definition}
The {\it distance} along edge $uv$ is
\[
d(uv) \;=\; \frac{1}{w(u,v)}.
\]
\end{definition}

\begin{definition}
For a graph $G=(V,E)$ define the following:
\begin{enumerate}
    \item $N_n(v)$ denotes the set of $n$-order neighbors of $v$;
    \item $P(u,v)$ is the set of all paths from $u$ to $v$;
    \item for a path $p:(u=v_0,\dots,v_n=v)\in P(u,v)$,
          \[
          a(p)=\sum_{i=0}^{n-1} d(v_i,v_{i+1}), \qquad 
          b(p)=\sum_{i=0}^{n-1} w(v_i,v_{i+1});
          \]
          and we set $D(u,v)=\min_{p\in P(u,v)} a(p)$;
    \item $k_u = |N_1(u)|$ is the (unweighted) degree of $u$ and
          $s_u = \sum_{x\in N(u)} w(u,x)$ its weighted degree (strength).
\end{enumerate}
\end{definition}

Link prediction is a field that seeks to identify potential future edges in a graph by considering properties of the current graph. A link-prediction score $s(u,v)$ is computed for each non-adjacent pair $u,v\in V$.  We focus on low-cost local measures that consider only first‐ and second‐order neighborhoods:

\begin{itemize}
\item \textbf{Common Neighbors (CN)}~\cite{commonneighbors}: 
      \[
      s^{CN}_{u,v}=|N(u)\cap N(v)|.
      \]

\item \textbf{Resource Allocation (RA)}~\cite{resourceallocation}:
      \[
      s^{RA}_{u,v}= \sum_{x\in N(u)\cap N(v)} \frac{1}{|N(x)|}.
      \]

\item \textbf{Novel Resource Allocation Varient RA–2}:  

 This RA$-2$ is proposed in \textit{The Social Sphere Model} \cite{lin2025socialspheremodelheuristic}. this involves approximating the probability that a common neighbor $x$ links two of its friends by $\tfrac{2}{|N(x)|^2}$.  We define
      \[
      s^{RA2}_{u,v}:= \sum_{x\in N(u)\cap N(v)} \frac{2}{|N(x)|^{2}}.
      \]
      Weighted analogues replace $|N(x)|$ with $s_x$.
\end{itemize}

Other path-based variants such as Local Path, Jaccard, and quasi-local extensions exist; however we omit formal definitions here for brevity. These definitions supply all notation required for our empirical study; full derivations and additional similarity metrics appear as defined in \cite{lin2025socialspheremodelheuristic}.

\subsection{The Social Sphere Model}
\label{sec:ssm}

In what follows we shall  briefly summarize the \textit{Social Sphere Model} introduced in~\cite{lin2025socialspheremodelheuristic}, a heuristic framework for identifying future influencers in evolving networks. The model combines local link prediction techniques with centrality-based node selection, offering an interpretable and computationally efficient alternative to more complex influence prediction algorithms.

The approach proceeds in two main steps. First, link prediction scores are computed between non-adjacent node pairs using similarity metrics such as CN, RA, or RA-2. These scores are normalized and interpreted as probabilities of future connection. To account for a time horizon of \( t \) steps into the future, the probability that an edge will have formed between nodes \( u \) and \( v \) is modeled as
\[
w_t(u,v) = 1 - (1 - p_{u,v})^t,
\]
where \( p_{u,v} \) is the initial similarity score. These adjusted weights define a predicted graph capturing expected future connectivity.

In the second step, a centrality measure (e.g., weighted degree) is applied to the predicted graph to identify top-\( k \) influential nodes. These are interpreted as the anticipated influencers at time \( t \), accounting for both evolving structure and anticipated ties.

The full model pipeline is described in Algorithm~\ref{alg:futureinfluencers}. Further theoretical discussion and design motivation can be found in~\cite{lin2025socialspheremodelheuristic}. In this paper, we focus on empirically evaluating the model across varying network densities, prediction horizons, and similarity metrics. In particular, we assess its ability to detect \textit{latent influencers}—nodes that are not prominent in the initial network but emerge as central over time.

\begin{algorithm}
\caption{Social Sphere Future Influencer Prediction}\label{alg:futureinfluencers}
\begin{algorithmic}
\Require{Graph $G$, Similarity Metric $M$, Top-$k$ Algorithm $c$, Time Horizon $t$}
\State Compute similarity scores $s^{M}_{u,v}$ for all non-adjacent pairs $(u,v)$.
\For{each $(u,v)$ with $s^{M}_{u,v} > 0$}
    \State Add edge $uv$ with weight $w_t(u,v) = 1 - (1 - s^{M}_{u,v})^t$
    \State Assign distance $d_{uv} = 1 / w_t(u,v)$
\EndFor
\State Apply $c$ to identify top-$k$ central nodes in the predicted graph
\end{algorithmic}
\end{algorithm}

\section{Methods}
\label{sec:methods}

To evaluate the Social Sphere Model in a practical setting, we conduct a series of experiments on a real-world temporal network. Our goal is to assess the model’s ability to identify future influencers under varying network densities and prediction horizons.

Given a graph $G$, we generate a training graph $G_{\text{train}}$ by randomly sampling a fraction of its edges. Using $G_{\text{train}}$, we construct a predicted future graph following the procedure described in~\cite{lin2025socialspheremodelheuristic}, and then apply top-$k$ centrality-based algorithms to identify influential nodes. The predicted influencers are subsequently evaluated on the full graph $G$ using contagion-based influence spread models.
Each configuration (i.e., choice of edge sampling rate, similarity metric, and time horizon) is tested in ten independent trials to account for stochastic variation, and results are averaged across runs.

\subsection{Evaluation Metrics}
\label{subsec:evalmethods}

We assess model performance using two main criteria:
\begin{itemize}
    \item \textbf{Influence Spread:} The proportion of nodes reached over time when predicted influencers are used as initial spreaders in $G$.
    \item \textbf{Prediction Accuracy:} Alignment between predicted and actual influencers, measured by:
    \begin{enumerate}
        \item \emph{Influencer Overlap:} The fraction of predicted top-$k$ influencers that also appear in the true top-$k$ set from $G$.
        \item \emph{Mean Squared Error (MSE):} Discrepancy in influence dynamics:
        \[
        \text{MSE} = \frac{1}{r} \sum_{t=1}^r \bigl(O(t) - P(t)\bigr)^2,
        \]
        where $P(t)$ is the fraction of nodes influenced at time $t$ by predicted influencers, and $O(t)$ the same using ground-truth influencers.
    \end{enumerate}
\end{itemize}

\subsection{Contagion Models}
\label{subsec:contagionmodels}

\begin{figure}[htb]
    \centering
    \begin{subfigure}[b]{0.15\textwidth}
        \centering
        \includegraphics[width=\textwidth]{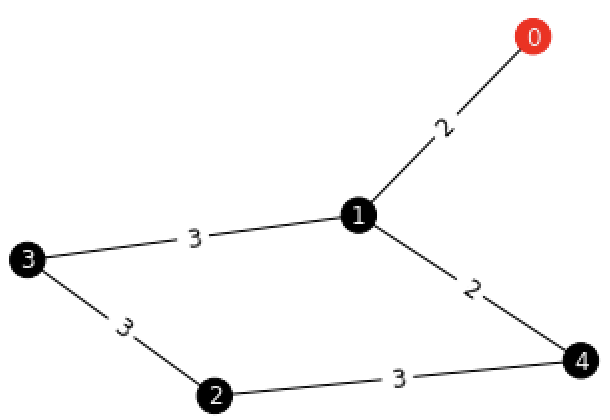}
        \caption{Simple Contagion at time $t=0$}
        \label{fig:sc0}
    \end{subfigure}
    \hfill
    \begin{subfigure}[b]{0.15\textwidth}
        \centering
        \includegraphics[width=\textwidth]{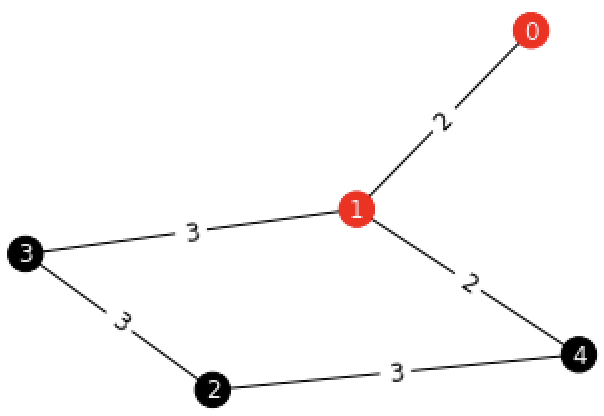}
        \caption{Simple Contagion at time $t=1$}
        \label{fig:sc1}
    \end{subfigure}
    \hfill
    \begin{subfigure}[b]{0.15\textwidth}
        \centering
        \includegraphics[width=\textwidth]{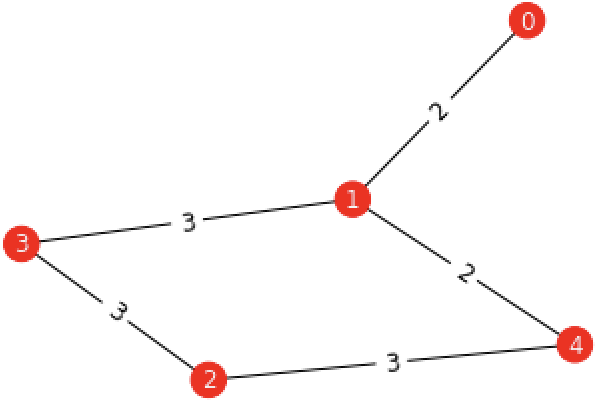}
        \caption{Simple Contagion at time $t=7$}
        \label{fig:sc2}
    \end{subfigure}

    \begin{subfigure}[b]{0.15\textwidth}
        \centering
        \includegraphics[width=\textwidth]{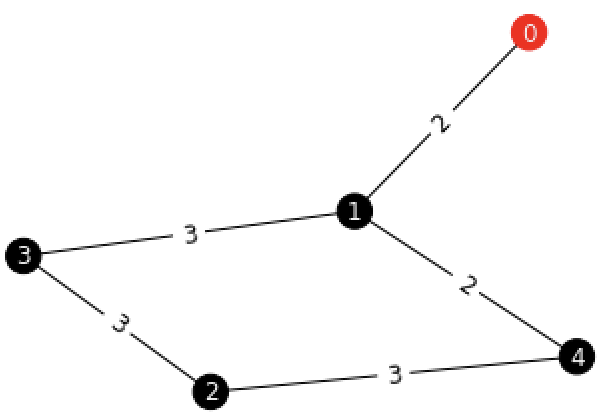}
        \caption{Complex Contagion at time $t=0$}
        \label{fig:cc0}
    \end{subfigure}
    \hfill
    \begin{subfigure}[b]{0.15\textwidth}
        \centering
        \includegraphics[width=\textwidth]{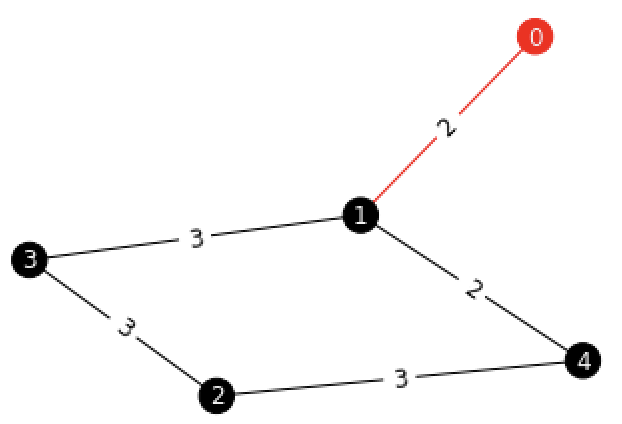}
        \caption{Complex Contagion at time $t=2$}
        \label{fig:cc1}
    \end{subfigure}
    \hfill
    \begin{subfigure}[b]{0.15\textwidth}
        \centering
        \includegraphics[width=\textwidth]{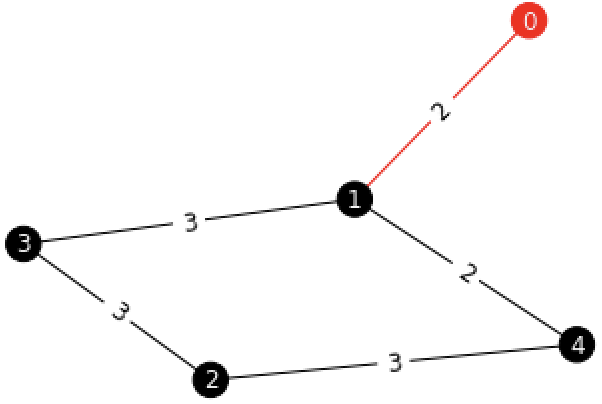}
        \caption{Complex Contagion at time $t=7$}
        \label{fig:cc2}
    \end{subfigure}

    \caption{Contagion examples with initial infected node 0 and distances marked on edges (successful infections and successful interactions for complex contagion are marked with red) \cite{lin2025socialspheremodelheuristic}.}
    \label{fig:contagionex}
\end{figure}

We simulate influence spread using the following two stylized models (Examples of each contagion model are illustrated in Figure~\ref{fig:contagionex}):

\paragraph{Simple Contagion:} A node becomes influenced if there exists a path from an initial spreader whose total edge distance is less than or equal to time $t$.
 The infected set at time $t$ is:
\[
I(t) = \{v \in V \mid \exists u \in I(0): D(u,v) \le t \}.
\]

\paragraph{Complex Contagion:} A node is influenced (or infected) if the sum of incoming influence weights from currently influenced neighbors exceeds a threshold.
 
The infected set at time $t$ is:
\[
I(t) = \left\{v \in V \,\middle|\, \sum_{u \in I(t-1)} w(u,v) \ge \theta \cdot s_v \right\},
\]
where $\theta \in [0,1]$ and $s_v$ is the strength of node $v$.

\section{SSM for empirical networks}
\label{sec:results}

\label{subsec:experimentalsetup}
In what follows we shall apply the Social Sphere Model to the arXiv General Relativity and Quantum Cosmology (GR-QC) collaboration network~\cite{cagrqc,snapnets}, which includes 5,242 nodes and 14,496 edges. Influencers are identified using a top-$k$ selection with $k = 25$, and we also evaluate single-node influence for reference. To assess the model under different temporal and structural conditions, we consider two experimental scenarios:

\begin{itemize}
    \item \textbf{Near-term prediction:} 90\% of edges sampled, with a prediction horizon of $t = 1$.
    \item \textbf{Long-term prediction:} 70\% of edges sampled, with a prediction horizon of $t = 3$.
\end{itemize}

These two setups allow us to evaluate the model’s performance under both high-information (dense) and low-information (sparse) settings. In each case, the training graph is generated by sampling the stated fraction of edges from the full network. The predicted future graph is then constructed following the Social Sphere Model framework, and influencer performance is evaluated using both simple and complex contagion models on the full graph.
Our analysis of the GR-QC dataset reveals several key findings:

\begin{itemize}
    \item The RA–2 similarity metric achieves the lowest average mean squared error (MSE) across trials.
    \item In some cases, the model identifies \emph{latent influencers}—nodes that outperform the actual top influencers when evaluated on the original graph.
    \item Adjusted versions of standard centrality algorithms, when applied within the Social Sphere Model, consistently outperform their unmodified counterparts.
\end{itemize}

\begin{table*}
    \centering
    \begin{tabular}{ | p{2.3cm} | p{1.7cm} | p{1.5cm} | p{2cm} | p{2cm} | p{2cm} | p{1.5cm} | p{2cm} | p{1.5cm}|}
        \hline
        \textbf{Centrality Metrics} & \textbf{Common Neighbors} & \textbf{Jaccard} & \textbf{Local Path} & \textbf{Quasi-local RA-2} & \textbf{Quasi-local RA} & \textbf{RA-2} & \textbf{Resource Allocation} & \textbf{Overall}\\
        \hline
        Balanced Index & 0.0151 & 0.0106 & 0.0292 & 0.0260 & 0.0247 & \textbf{0.0104} & 0.0116 & 0.0182 \\
        Betweenness & 0.0110 & 0.0094 & 0.0224 & 0.0205 & 0.0194 & 0.0087 & \textbf{0.0092} & 0.0144 \\
        Closeness & 0.0170 & 0.0138 & 0.0295 & 0.0262 & 0.0249 & \textbf{0.0133} & 0.0135 & 0.0197 \\
        ClusterRank & 0.0174 & 0.0151 & 0.0301 & 0.0296 & 0.0287 & \textbf{0.0146} & 0.0149 & 0.0215 \\
        Complex Path Centrality & 0.0475 & 0.0480 & 0.0442 & 0.0434 & \textbf{0.0431} & 0.0486 & 0.0484 & 0.0462 \\
        Degree & 0.0107 & 0.0097 & 0.0229 & 0.0219 & 0.0195 & 0.0086 & \textbf{0.0084} & 0.0145 \\
        Eigenvector & 0.0109 & 0.0095 & 0.0240 & 0.0223 & 0.0212 & \textbf{0.0091} & 0.0098 & 0.0152 \\
        H-index & 0.0156 & 0.0152 & 0.0318 & 0.0282 & 0.0268 & \textbf{0.0150} & 0.0152 & 0.0211 \\
        k-core & 0.0117 & \textbf{0.0103} & 0.0222 & 0.0189 & 0.0187 & 0.0109 & 0.0107 & 0.0148 \\
        LeaderRank & 0.0108 & 0.0083 & 0.0229 & 0.0211 & 0.0196 & \textbf{0.0076} & 0.0080 & 0.0140 \\
        LocalRank & 0.0094 & 0.0080 & 0.0231 & 0.0216 & 0.0205 & \textbf{0.0077} & 0.0079 & 0.0140 \\
        PageRank & 0.0130 & \textbf{0.0087} & 0.0266 & 0.0240 & 0.0237 & 0.0099 & 0.0094 & 0.0165 \\
        \hline
        Overall & 0.0158 & 0.0139 & 0.0274 & 0.0253 & 0.0242 & \textbf{0.0137} & 0.0139 & 0.0192 \\
        \hline
    \end{tabular}
    \caption{Average MSE for centrality metrics over all algorithms and contagion models on $70\%$ training graphs.}
    \label{tab:cagrqc_70_mse_on_metric}
\end{table*}

\begin{table*}
    \centering
    \begin{tabular}{ | p{2.3cm} | p{1.7cm} | p{1.5cm} | p{2cm} | p{2cm} | p{2cm} | p{1.5cm} | p{2cm} | p{1.5cm}|}
        \hline
        \textbf{Centrality Metrics} & \textbf{Common Neighbors} & \textbf{Jaccard} & \textbf{Local Path} & \textbf{Quasi-local RA-2} & \textbf{Quasi-local RA} & \textbf{RA-2} & \textbf{Resource Allocation} & \textbf{Overall} \\
        \hline
        Balanced Index & 0.0132 & 0.0132 & 0.0221 & 0.0221 & 0.0218 & \textbf{0.0129} & 0.0132 & 0.0169 \\
        Betweenness & 0.0114 & 0.0107 & 0.0199 & 0.0192 & 0.0198 & \textbf{0.0102} & 0.0111 & 0.0146 \\
        Closeness & 0.0123 & 0.0117 & 0.0201 & 0.0198 & 0.0197 & \textbf{0.0111} & 0.0116 & 0.0152 \\
        ClusterRank & 0.0120 & 0.0120 & 0.0206 & 0.0202 & 0.0204 & 0.0120 & \textbf{0.0118} & 0.0156 \\
        Complex Path Centrality & 0.0732 & 0.0732 & 0.0376 & \textbf{0.0374} & 0.0377 & 0.0733 & 0.0734 & 0.0580 \\
        Degree & 0.0107 & 0.0102 & 0.0172 & 0.0169 & 0.0170 & \textbf{0.0099} & 0.0101 & 0.0131 \\
        Eigenvector & 0.0077 & 0.0074 & 0.0120 & 0.0117 & 0.0116 & \textbf{0.0069} & 0.0072 & 0.0092 \\
        H-index & 0.0125 & 0.0126 & 0.0182 & 0.0179 & 0.0177 & \textbf{0.0122} & 0.0124 & 0.0148 \\
        k-core & 0.0030 & \textbf{0.0029} & 0.0048 & 0.0045 & 0.0046 & 0.0031 & \textbf{0.0029} & 0.0037 \\
        LeaderRank & 0.0115 & 0.0108 & 0.0179 & 0.0175 & 0.0181 & \textbf{0.0104} & 0.0105 & 0.0138 \\
        LocalRank & 0.0107 & 0.0109 & 0.0198 & 0.0193 & 0.0195 & \textbf{0.0103} & 0.0105 & 0.0144 \\
        PageRank & 0.0122 & 0.0116 & 0.0204 & 0.0199 & 0.0199 & \textbf{0.0113} & 0.0117 & 0.0153 \\
        \hline
        Overall & 0.0159 & 0.0156 & 0.0192 & 0.0189 & 0.0190 & \textbf{0.0153} & 0.0155 & 0.0170 \\
        \hline
    \end{tabular}
    \caption{Average MSE for  centrality metrics over all algorithms and contagion models on $90\%$ training graphs.}
    \label{tab:cagrqc_90_mse_on_metric}
\end{table*}

\begin{table*}[htp]
    \centering
        \begin{tabular}{ | p{2cm} | p{1.7cm} |p{1.2cm} |p{1.7cm} |p{1.5cm} |p{1.7cm} |p{1.2cm} |p{1.7cm} |p{1.2cm} |}
        \hline
        Centrality Metric & Common Neighbors & Jaccard & Local Path & Quasi-Local RA & Quasi-Local RA-2 & RA-2 & Resource Allocation & Overall \\
        \hline
        Balanced Index & 0.3217 & \textbf{0.4166} & 0.3091 & 0.3714 & 0.3691 & 0.4086 & 0.4046 & 0.3716 \\
        Betweenness & 0.3811 & 0.3903 & 0.3549 & 0.3537 & 0.3503 & 0.3937 & \textbf{0.3943} & 0.3740 \\
        Closeness & 0.2897 & 0.3371 & 0.2720 & 0.3120 & 0.3154 & \textbf{0.3463} & 0.3411 & 0.3162 \\
        ClusterRank & \textbf{0.2794} & 0.2731 & 0.2217 & 0.1891 & 0.1806 & 0.2760 & 0.2714 & 0.2416 \\
        Complex Path Centrality & \textbf{0.2171} & 0.2160 & 0.1326 & 0.1320 & 0.1343 & \textbf{0.2171} & 0.2166 & 0.1808 \\
        Degree & \textbf{0.3634} & 0.3520 & 0.3400 & 0.3211 & 0.3211 & 0.3594 & 0.3594 & 0.3452 \\
        Eigenvector & \textbf{0.2691} & 0.2154 & 0.2577 & 0.2091 & 0.2057 & 0.2240 & 0.2189 & 0.2286 \\
        H-index & \textbf{0.1846} & 0.1509 & 0.0617 & 0.0571 & 0.0560 & 0.1509 & 0.1497 & 0.1158 \\
        k-core & \textbf{0.4680} & 0.4623 & 0.4417 & 0.4394 & 0.4371 & 0.4634 & 0.4611 & 0.4533 \\
        LeaderRank & 0.3514 & 0.3491 & 0.3297 & 0.3177 & 0.3154 & \textbf{0.3560} & 0.3531 & 0.3389 \\
        LocalRank & \textbf{0.3000} & 0.2886 & 0.2200 & 0.2057 & 0.2017 & 0.2977 & \textbf{0.3000} & 0.2591 \\
        PageRank & \textbf{0.3766} & 0.3651 & 0.3503 & 0.3291 & 0.3263 & 0.3674 & 0.3674 & 0.3546 \\
        \hline
        Overall & 0.3169 & 0.3180 & 0.2743 & 0.2698 & 0.2678 & \textbf{0.3217} & 0.3198 & 0.2983 \\
        \hline
    \end{tabular}
    \caption{Average overall accuracies for centrality metrics on $70\%$ training graphs  from the arXiv GrQc dataset.}
    \label{tab:cagrqc_70_sim_on_metric}
\end{table*}

\begin{table*}[htp]
    \centering
        \begin{tabular}{ | p{2cm} | p{1.7cm} |p{1.7cm} |p{1.5cm} |p{1.5cm} |p{1.7cm} |p{1.2cm} |p{1.7cm} |p{1.2cm} |}
        \hline
        Centrality Metric & Common Neighbors & Jaccard & Local Path & Quasi-Local RA & Quasi-Local RA-2 & RA-2 & Resource Allocation & Overall \\
        \hline
        Balanced Index & 0.4166 & \textbf{0.4869} & 0.3989 & 0.4589 & 0.4571 & 0.4857 & 0.4851 & 0.4556 \\
        Betweenness & \textbf{0.4743} & 0.4703 & 0.4366 & 0.4343 & 0.4349 & 0.4709 & 0.4697 & 0.4558 \\
        Closeness & 0.3874 & 0.3983 & 0.3714 & 0.3857 & 0.3846 & \textbf{0.4006} & 0.4000 & 0.3897 \\
        ClusterRank & \textbf{0.2937} & 0.2743 & 0.2080 & 0.1960 & 0.1903 & 0.2743 & 0.2771 & 0.2448 \\
        Complex Path Centrality & 0.1680 & 0.1640 & 0.1560 & 0.1566 & 0.1571 & \textbf{0.1720} & 0.1680 & 0.1631 \\
        Degree & \textbf{0.4606} & 0.4206 & 0.4229 & 0.3869 & 0.3869 & 0.4223 & 0.4194 & 0.4171 \\
        Eigenvector & \textbf{0.3789} & 0.3131 & 0.3600 & 0.2977 & 0.3029 & 0.3217 & 0.3189 & 0.3276 \\
        H-index & \textbf{0.1834} & 0.1743 & 0.0863 & 0.0869 & 0.0834 & 0.1714 & 0.1726 & 0.1369 \\
        k-core & \textbf{0.5274} & 0.5217 & 0.4909 & 0.4829 & 0.4857 & 0.5211 & 0.5211 & 0.5073 \\
        LeaderRank & \textbf{0.4640} & 0.4326 & 0.4274 & 0.4006 & 0.4006 & 0.4360 & 0.4349 & 0.4280 \\
        LocalRank & \textbf{0.3280} & 0.3194 & 0.2246 & 0.2200 & 0.2183 & 0.3194 & 0.3194 & 0.2784 \\
        PageRank & \textbf{0.4851} & 0.4480 & 0.4497 & 0.4143 & 0.4177 & 0.4509 & 0.4486 & 0.4449 \\
        \hline
        Overall & \textbf{0.3806} & 0.3686 & 0.3360 & 0.3267 & 0.3266 & 0.3705 & 0.3696 & 0.3541 \\
        \hline
    \end{tabular}
    \caption{Average overall accuracies for centrality metrics on $90\%$ training graphs  from the arXiv GrQc dataset.}
    \label{tab:cagrqc_90_sim_on_metric}
\end{table*}
\subsection{Mean Squared Error (MSE) Comparison}

We evaluated the predictive performance of various link prediction metrics paired with centrality-based influencer selection by computing the mean squared error (MSE) between predicted and actual influence curves. Tables~\ref{tab:cagrqc_sms_mse_70} and~\ref{tab:cagrqc_sms_mse_90} report the average MSE across all top-$k$ algorithms, under both sparse ($70\%$ edges) and dense ($90\%$ edges) training scenarios.

In both settings, the RA-2 similarity metric consistently achieves the lowest overall MSE, confirming its robustness to different graph densities. For instance:
\begin{itemize}
    \item At $70\%$ edge sampling (Table~\ref{tab:cagrqc_sms_mse_70}), RA-2 has the lowest overall MSE ($0.0144$), including the lowest score in the complex contagion model ($0.0225$).
    \item At $90\%$ edge sampling (Table~\ref{tab:cagrqc_sms_mse_90}), RA-2 again yields the best overall MSE ($0.0159$), and remains top-ranked under the complex contagion model.
\end{itemize}

While Jaccard performs slightly better in the simple contagion setting at $70\%$ sampling ($0.0049$), it is consistently outperformed by RA-2 under complex contagion, which more accurately reflects information spread requiring social reinforcement. This suggests that RA-2’s incorporation of second-order neighborhood structure and degree normalization makes it better suited for capturing subtle influence dynamics, especially when training data is limited.

\begin{table}[H]
\centering
    \begin{tabular}{lccc}
        \hline
        \textbf{Prediction Metric} & \textbf{Complex} & \textbf{Simple} & \textbf{Overall} \\
        \hline
        Common Neighbors & 0.02641 & 0.00557 & 0.01599 \\
        Jaccard & 0.02282 & \textbf{0.00493} & \textbf{0.01388} \\
        Local Path & 0.04713 & 0.00811 & 0.02762 \\
        Quasi-Local RA & 0.04289 & 0.00753 & 0.02521 \\
        Quasi-Local RA-2 & 0.04081 & 0.00738 & 0.02410 \\
        RA-2 & \textbf{0.02254} & 0.00635 & 0.01444 \\
        Resource Allocation & 0.02278 & 0.00543 & 0.01411 \\
        \hline
        Overall & 0.03220 & 0.00647 & 0.01933 \\
        \hline
    \end{tabular}
    \caption{Average MSE over top-$k$ algorithms for link prediction metrics using $70\%$ training graphs from the arXiv GR-QC dataset~\cite{cagrqc}.}
    \label{tab:cagrqc_sms_mse_70}
\end{table}

\begin{table}[H]
\centering
    \begin{tabular}{lccc}
        \hline
        \textbf{Prediction Metric} & \textbf{Complex} & \textbf{Simple} & \textbf{Overall} \\
        \hline
        Common Neighbors & 0.01914 & 0.01368 & 0.01641 \\
        Jaccard & 0.01854 & \textbf{0.01334} & 0.01594 \\
        Local Path & 0.02607 & 0.01365 & 0.01986 \\
        Quasi-Local RA & 0.02539 & 0.01392 & 0.01966 \\
        Quasi-Local RA-2 & 0.02558 & 0.01371 & 0.01965 \\
        RA-2 & \textbf{0.01804} & 0.01372 & \textbf{0.01588} \\
        Resource Allocation & 0.01846 & 0.01414 & 0.01630 \\
        \hline
        Overall & 0.02160 & 0.01374 & 0.01767 \\
        \hline
    \end{tabular}
    \caption{Average MSE over top-$k$ algorithms for link prediction metrics using $90\%$ training graphs from the arXiv GR-QC dataset~\cite{cagrqc}.}
    \label{tab:cagrqc_sms_mse_90}
\end{table}

These results suggest that RA-2 offers strong predictive performance across a variety of conditions, balancing both accuracy and generalizability. Its effectiveness is especially evident in the complex contagion setting, which mimics more realistic patterns of influence in social systems.

\subsection{Accuracy Comparison}

To complement the MSE analysis, we assess the accuracy of influencer prediction by comparing the overlap between predicted top-$k$ influencers and the actual influencers in the full graph. Tables~\ref{tab:cagrqc_70_sim_on_metric} and~\ref{tab:cagrqc_90_sim_on_metric} report these results under the 70\% and 90\% training edge regimes, respectively.

As expected, accuracy scores are generally higher under the 90\% setting, where the predicted graph more closely resembles the true network. In the sparser 70\% case, more variation is observed due to the increased uncertainty in link prediction.

\paragraph{RA-2 Performance.}  
The RA-2 metric performs best overall in the 70\% training case, achieving the highest accuracy (0.3217), and ranks second overall in the 90\% case (0.3705). Notably, RA-2 also leads for several centrality measures in the 70\% case (including LeaderRank (0.3560), PageRank (0.3674), and Eigenvector (0.2240)) demonstrating its ability to preserve influence-relevant structure even under data sparsity. In the 90\% case, RA-2 maintains strong performance, topping accuracy for Closeness (0.4006) and Complex Path (0.1720) centralities.

\paragraph{Comparative Metric Trends.}
Common Neighbors achieves the highest overall accuracy (0.3806) in the 90\% case, outperforming RA-2 by a small margin. However, RA-2 remains more consistent across both training regimes, highlighting its robustness. Jaccard performs well under certain centralities in the sparse setting, e.g., Balanced Index (0.4166), Closeness (0.3371), but its overall accuracy (0.3180) is slightly lower and less stable due to over-penalization in high-degree regions when data is limited.

This contrast between Jaccard and Common Neighbors reflects their structural differences. Jaccard normalizes for degree, which helps in dense graphs but hurts in sparse settings where node degrees are under-observed. For example, in the 70\% case, both have similar performance (Jaccard 0.3180 vs. Common Neighbors 0.3169), but in the 90\% case, Common Neighbors pulls ahead (0.3806 vs. 0.3686), particularly excelling on high-degree-biased centralities like k-core, Betweenness, Degree, and PageRank.

\paragraph{Underperforming Metrics.}
Local Path consistently ranks lowest in both regimes (overall: 0.2743 at 70\%, 0.3360 at 90\%), likely due to its reliance on global path information, which is difficult to reconstruct under edge sparsity. Complex Path Centrality also shows poor alignment, reinforcing that metrics requiring long-range information suffer under incomplete graphs.

\ 
Overall, RA-2 emerges as the most reliable metric across both data-rich and data-scarce settings, balancing performance, generalizability, and alignment with real influencers. Its second-order neighborhood modeling allows it to remain robust even when the input network is only partially observed.

\subsection{Influence Comparison:\\ Algorithm-Level Analysis}
\label{subsec:influence_algorithms}

We now examine the diffusion effectiveness of influencers selected by different algorithms under the Social Sphere Model. In particular, we evaluate how well these influencers propagate through the true future network, as measured by fraction infected over time in both simple and complex contagion settings.

\begin{figure*}[t]
    \centering
    \begin{subfigure}{0.42\textwidth}
        \includegraphics[width=\textwidth]{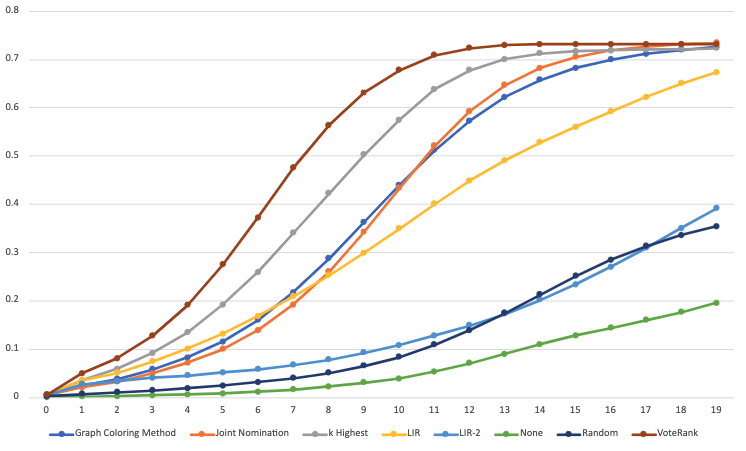}
        \caption{Complex contagion}
    \end{subfigure}
    \hfill
    \begin{subfigure}{0.42\textwidth}
        \includegraphics[width=\textwidth]{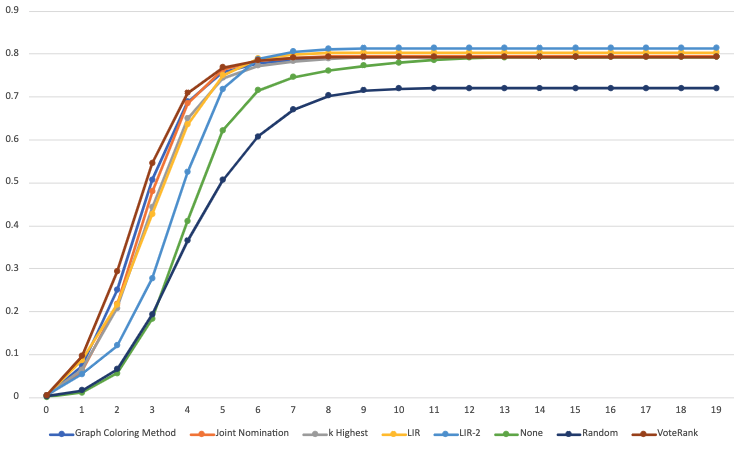}
        \caption{Simple contagion}
    \end{subfigure}
    \caption{Average performance of each algorithm over time in 70\% training graphs from the arXiv GR-QC dataset.}
    \label{cagrqc_70_alg_performance}
\end{figure*}

\begin{figure*}[t]
    \centering
    \begin{subfigure}{0.42\textwidth}
        \includegraphics[width=\textwidth]{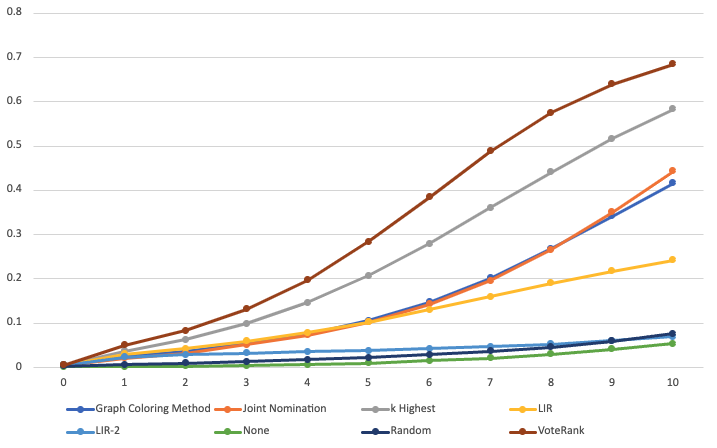}
        \caption{Complex contagion}
    \end{subfigure}
    \hfill
    \begin{subfigure}{0.42\textwidth}
        \includegraphics[width=\textwidth]{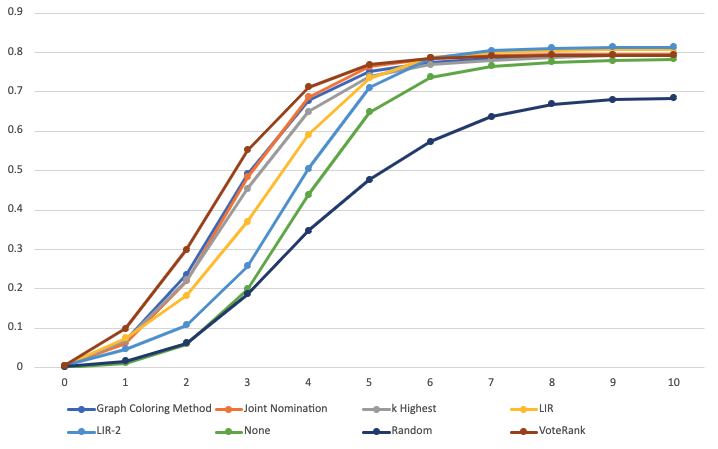}
        \caption{Simple contagion}
    \end{subfigure}
    \caption{Average performance of each algorithm over time in 90\% training graphs from the arXiv GR-QC dataset.}
    \label{cagrqc_90_alg_performance}
\end{figure*}

Figures~\ref{cagrqc_70_alg_performance} and~\ref{cagrqc_90_alg_performance} show the average fraction of nodes infected over time for each algorithm. Across all settings, \textit{VoteRank} is consistently the top-performing method, especially in simple contagion where it reaches 80\% infection within 7 time steps in the 70\% training case, and performs even faster in the 90\% case. 
In complex contagion scenarios, where reinforcement thresholds limit diffusion, VoteRank still leads in overall reach and speed. The $k$-Highest algorithm (light gray) also performs competitively, especially under complex contagion, indicating that in highly structured graphs, naive centrality-based seeding can be surprisingly effective.

Interestingly, whilst LIR and LIR-2 had strong performance in random networks, they struggled in this empirical dataset. The fragmented nature of the arXiv GR-QC network, with many weakly connected clusters, reduces the benefit of their spacing mechanisms. In contrast, Graph Coloring and Joint Nomination, which explicitly diversify seed placement, perform better in this setting.
Simple contagion leads to rapid saturation in all cases, while complex contagion highlights differences in algorithmic influence power. These trends emphasize the need to tailor influencer selection to both network structure and diffusion model.

 \subsection{Influence Comparison: \\Prediction Metrics and Centralities}
\label{subsec:influence_metrics}

\begin{figure*}[t]
    \centering
    \begin{subfigure}[b]{0.42\textwidth}
        \includegraphics[width=\textwidth]{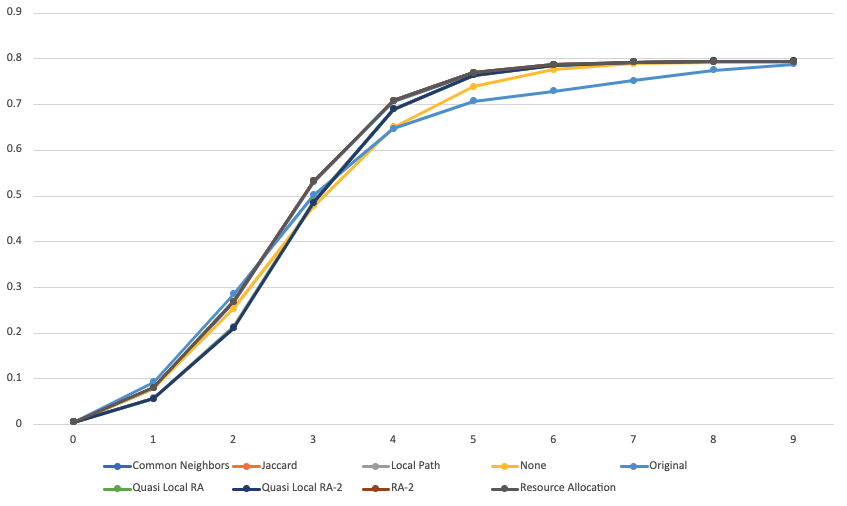}
        \caption{Graph Coloring}
    \end{subfigure}
    \hfill
    \begin{subfigure}[b]{0.42\textwidth}
        \includegraphics[width=\textwidth]{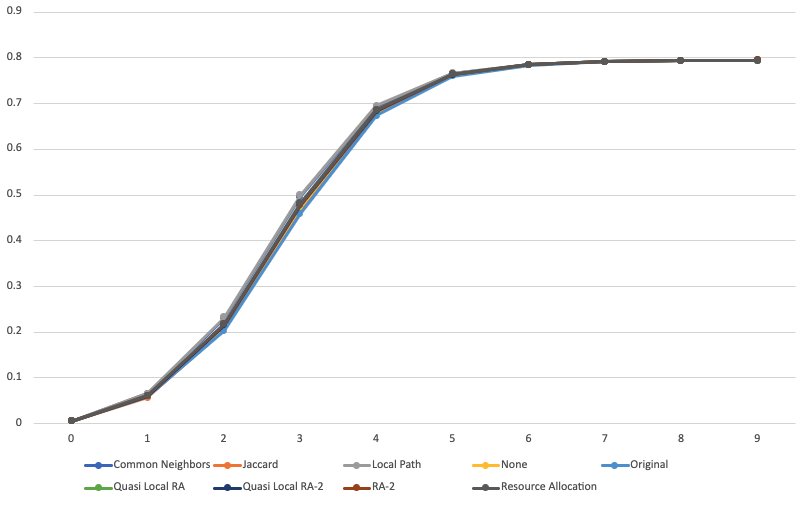}
        \caption{Joint Nomination}
    \end{subfigure}
    \hfill
    \begin{subfigure}[b]{0.42\textwidth}
        \includegraphics[width=\textwidth]{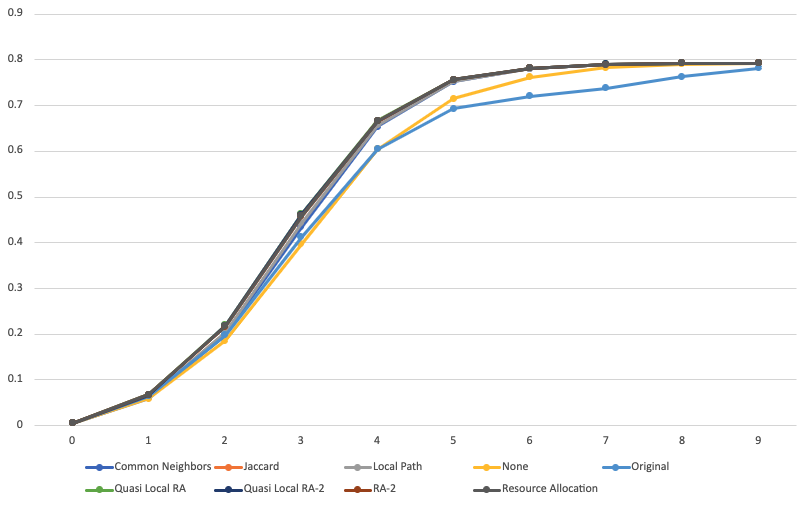}
        \caption{$k$-Highest}
    \end{subfigure}
    \hfill
    \begin{subfigure}[b]{0.42\textwidth}
        \includegraphics[width=\textwidth]{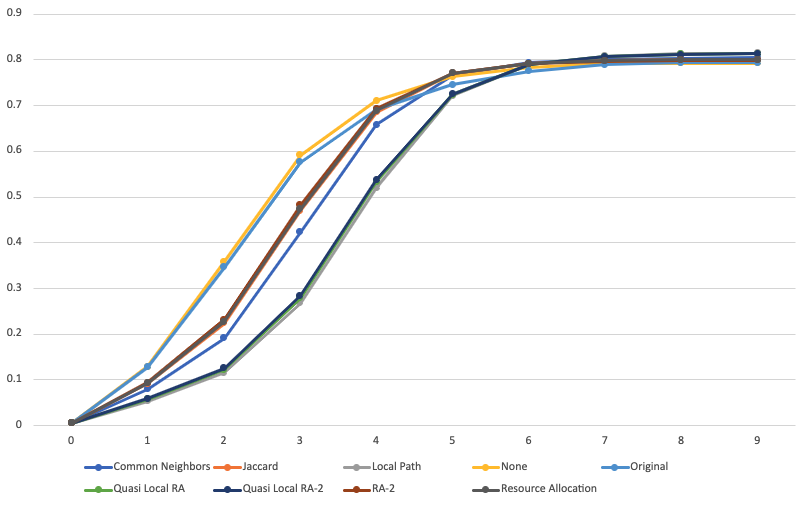}
        \caption{LIR}
    \end{subfigure}
    \hfill
    \begin{subfigure}[b]{0.42\textwidth}
        \includegraphics[width=\textwidth]{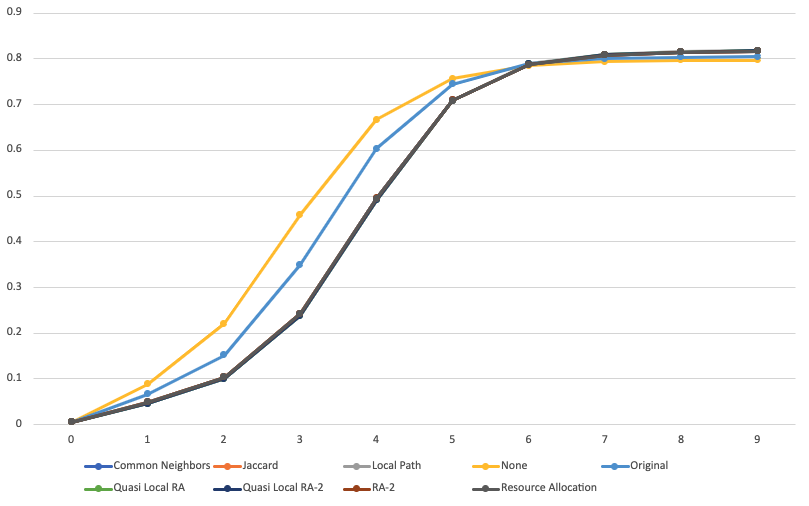}
        \caption{LIR-2}
    \end{subfigure}
    \hfill
    \begin{subfigure}[b]{0.42\textwidth}
        \includegraphics[width=\textwidth]{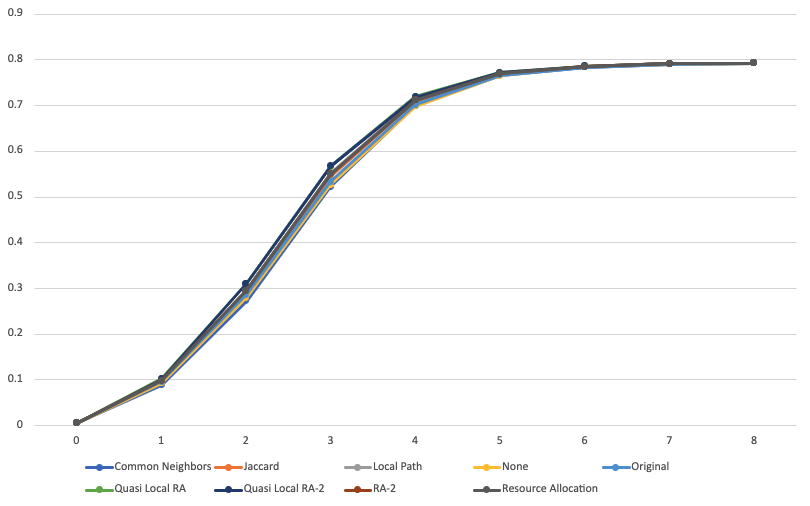}
        \caption{VoteRank}
    \end{subfigure}
    \caption{Fraction infected over time using different link prediction metrics in simple contagion (70\% training).}
    \label{fig:cagrqc_70_alg_sms_sc}
\end{figure*}

We now examine how the choice of link prediction metric and centrality affects influencer performance, holding the algorithm fixed. Figures~\ref{fig:cagrqc_70_alg_sms_cc} and~\ref{fig:cagrqc_70_alg_sms_sc} display the performance over time for each top-$k$ algorithm when applied to the predicted graphs under different link prediction metrics. From the above, several key insights emerge:

\begin{figure*}[t]
    \centering
    \begin{subfigure}[b]{0.42\textwidth}
        \includegraphics[width=\textwidth]{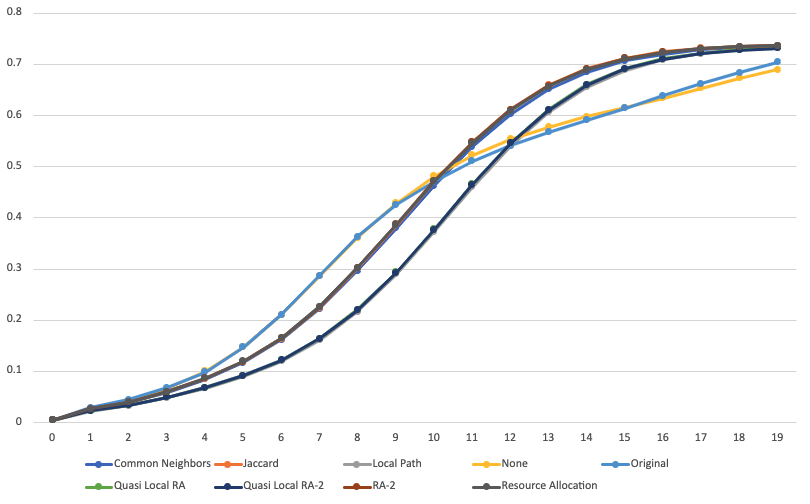}
        \caption{Graph Coloring}
    \end{subfigure}
    \hfill
    \begin{subfigure}[b]{0.42\textwidth}
        \includegraphics[width=\textwidth]{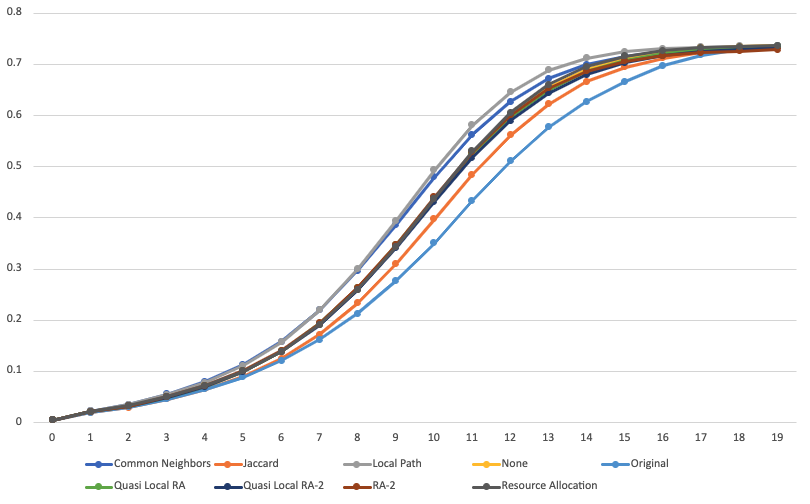}
        \caption{Joint Nomination}
    \end{subfigure}
    \hfill
    \begin{subfigure}[b]{0.42\textwidth}
        \includegraphics[width=\textwidth]{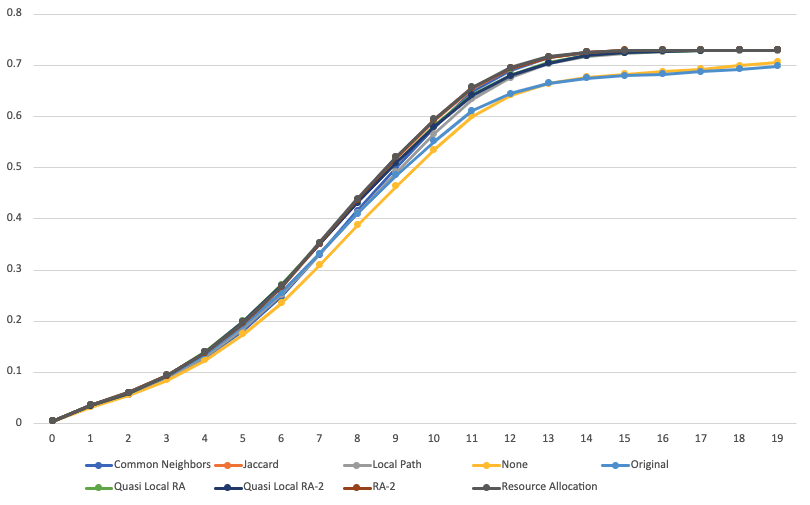}
        \caption{$k$-Highest}
    \end{subfigure}
    \hfill
    \begin{subfigure}[b]{0.42\textwidth}
        \includegraphics[width=\textwidth]{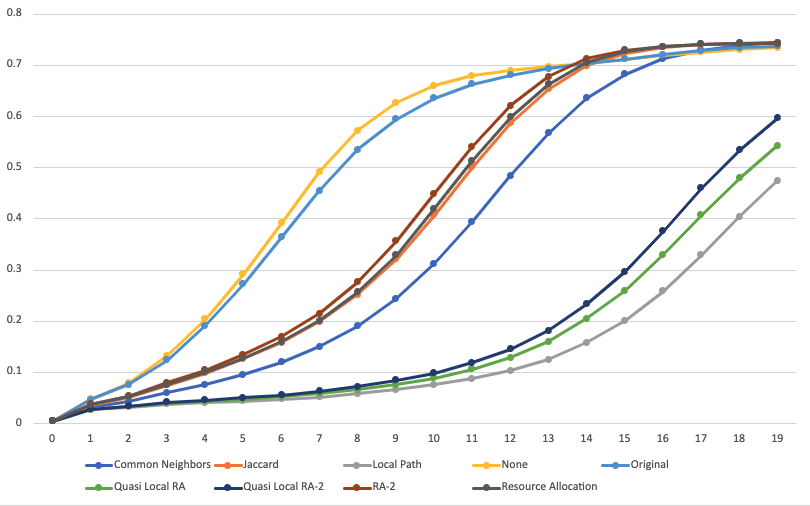}
        \caption{LIR}
    \end{subfigure}
    \hfill
    \begin{subfigure}[b]{0.42\textwidth}
        \includegraphics[width=\textwidth]{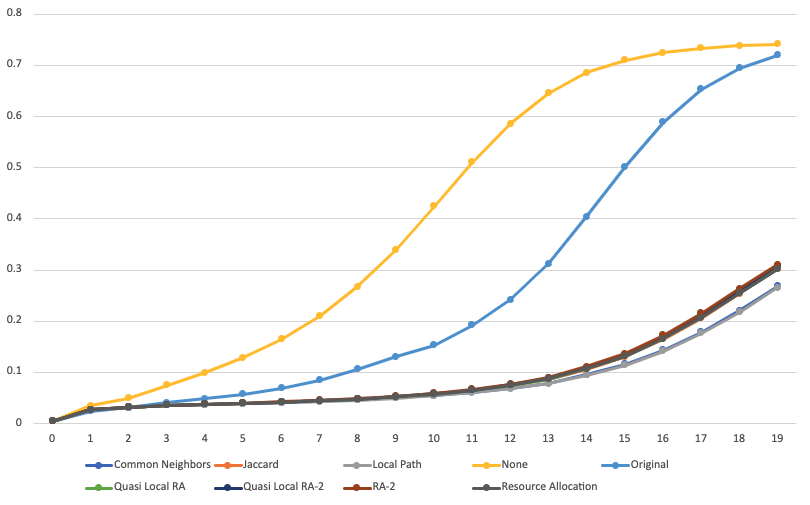}
        \caption{LIR-2}
    \end{subfigure}
    \hfill
    \begin{subfigure}[b]{0.42\textwidth}
        \includegraphics[width=\textwidth]{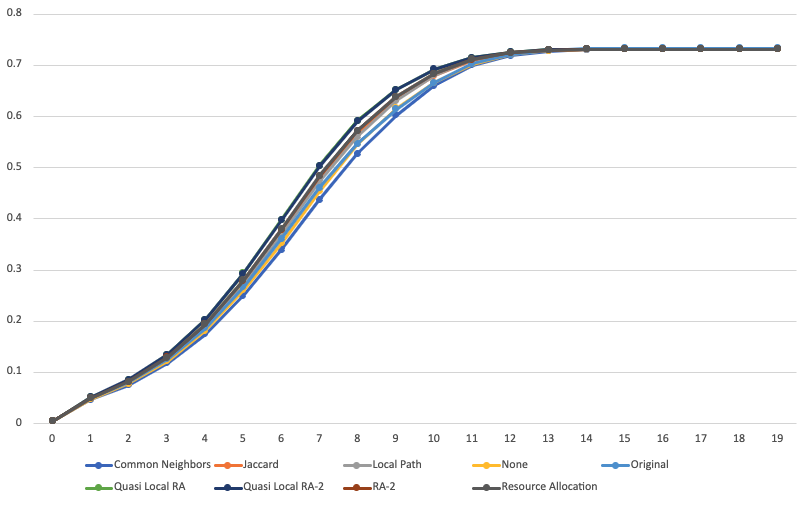}
        \caption{VoteRank}
    \end{subfigure}
    \caption{Fraction infected over time using different link prediction metrics in complex contagion (70\% training).}
    \label{fig:cagrqc_70_alg_sms_cc}
\end{figure*}

\begin{itemize}
    \item \textbf{Link prediction graphs can rival or exceed original graphs.} In Graph Coloring with simple contagion (Fig.~\ref{fig:cagrqc_70_alg_sms_sc}a), all metrics reach the same plateau as the original graph. Likewise, VoteRank results (Figs.~\ref{fig:cagrqc_70_alg_sms_cc}f and~\ref{fig:cagrqc_70_alg_sms_sc}f) show that Social Sphere-predicted graphs often converge to the same influence levels as the original by $t=12$ (complex) or $t=5$ (simple).
    
    \item \textbf{Predicted graphs may even outperform the true graph.} RA-2 consistently yields higher plateaus in Graph Coloring (Fig.~\ref{fig:cagrqc_70_alg_sms_cc}a) and $k$-Highest (Fig.~\ref{fig:cagrqc_70_alg_sms_cc}c), suggesting that it surfaces latent influencers not obvious from the partial graph.

    \item \textbf{Link prediction helps more under complex contagion.} In simple contagion, fast saturation leaves little room for improvement. However, under complex contagion, precision in seed placement—enabled by metrics like RA-2 and Jaccard—can drive significant gains.

    \item \textbf{Prediction metric choice matters.} RA-2, which penalizes high-degree neighbors, often produces better cascades than simpler metrics like Common Neighbors. Its deeper view of second-order neighborhoods appears to boost influence identification especially in sparse or incomplete networks.
\end{itemize}

In the next subsection, we further break down how each centrality behaves across different predicted graphs, providing evidence that degree-aware completions enable path-sensitive metrics (e.g., betweenness, closeness) to surface stronger influencers.

\subsection{Centrality Behavior Across Predicted Graphs}
\label{subsec:centrality_behavior}

\begin{figure*}[t]
    \centering
    \begin{subfigure}[b]{0.40\textwidth}
        \includegraphics[width=\textwidth]{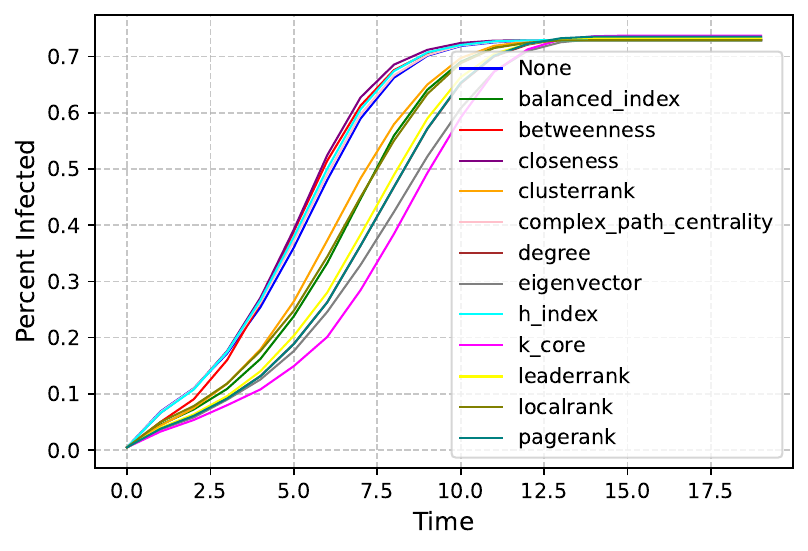}
        \caption{Local Path}
    \end{subfigure}
    \begin{subfigure}[b]{0.40\textwidth}
        \includegraphics[width=\textwidth]{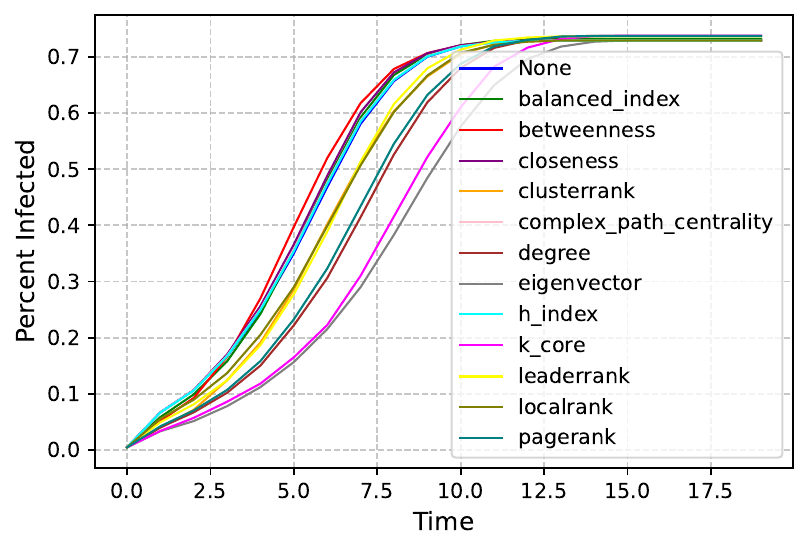}
        \caption{Quasi-Local RA-2}
    \end{subfigure}
    \begin{subfigure}[b]{0.40\textwidth}
        \includegraphics[width=\textwidth]{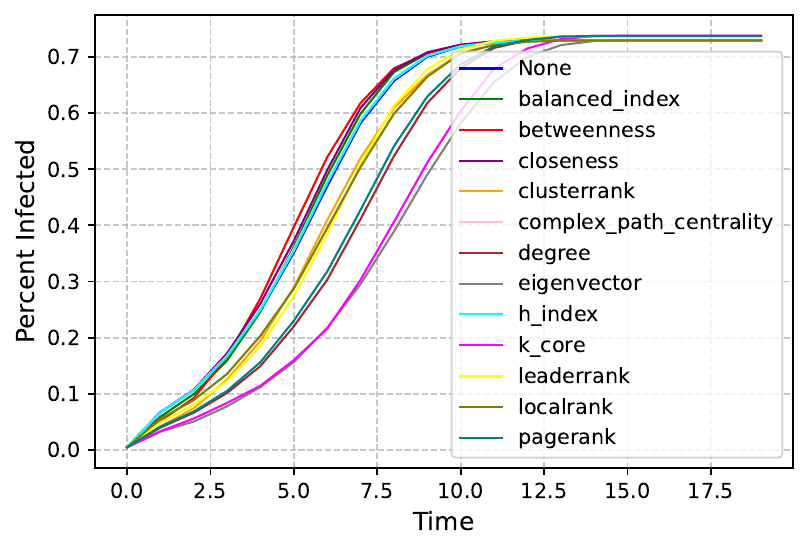}
        \caption{Quasi-Local RA}
    \end{subfigure}
    \begin{subfigure}[b]{0.40\textwidth}
        \includegraphics[width=\textwidth]{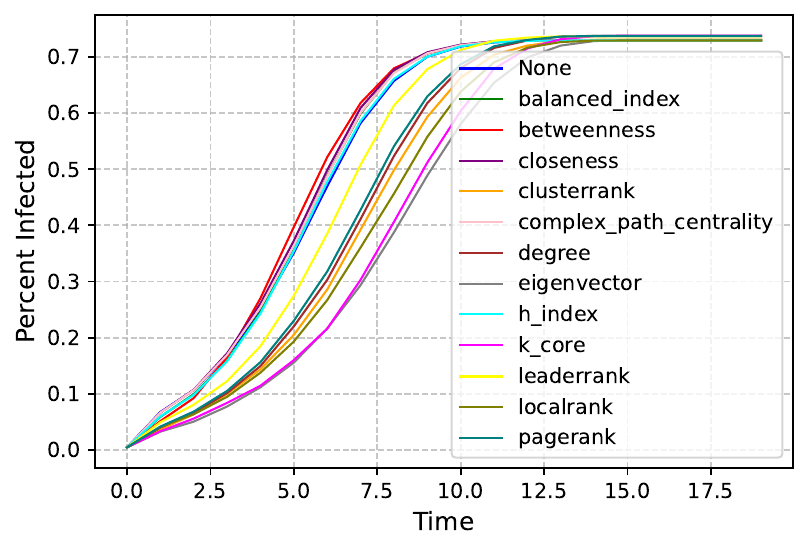}
        \caption{Resource Allocation}
    \end{subfigure}
    \caption{Centrality-VoteRank performance over predicted graphs for complex contagion (Part 2).}
    \label{fig:70_25_cagrqc_complex_part2}
\end{figure*}

We now investigate how different centrality scores perform when used as the selection basis for influencers on predicted graphs. Figures~\ref{fig:70_25_cagrqc_complex_part1} and~\ref{fig:70_25_cagrqc_complex_part2} illustrate the influence spread over time when VoteRank is modified to use different centrality rankings across a range of link-predicted completions. All experiments use 70\% training data and the complex contagion model. Across both figures, several patterns are evident:

\begin{figure*}[t]
    \centering
    \begin{subfigure}[b]{0.40\textwidth}
        \includegraphics[width=\textwidth]{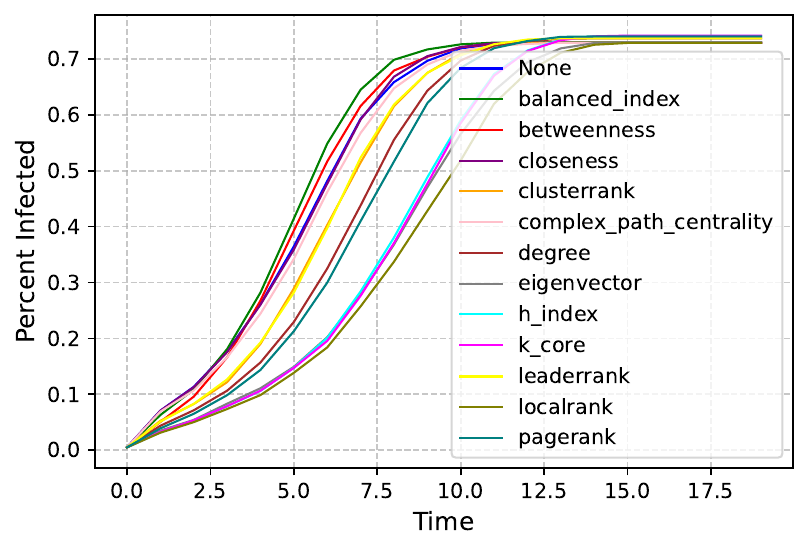}
        \caption{Original Graph}
    \end{subfigure}
    \begin{subfigure}[b]{0.40\textwidth}
        \includegraphics[width=\textwidth]{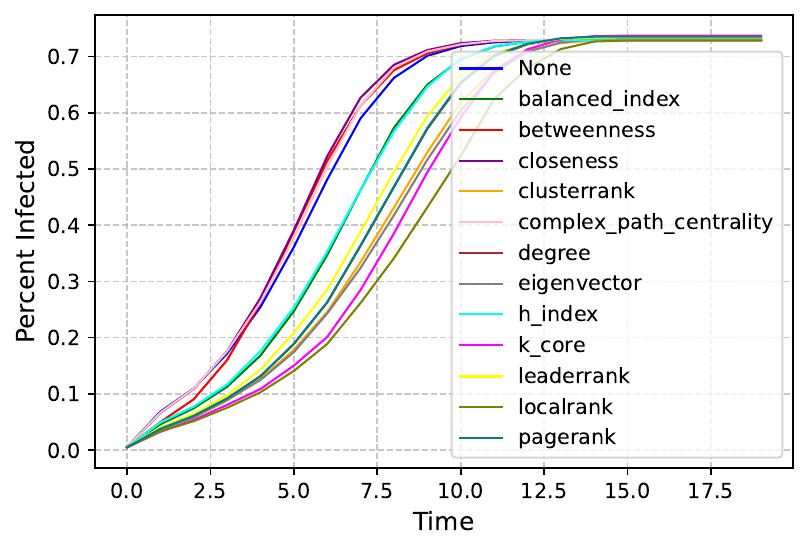}
        \caption{Common Neighbors}
    \end{subfigure}
    \begin{subfigure}[b]{0.40\textwidth}
        \includegraphics[width=\textwidth]{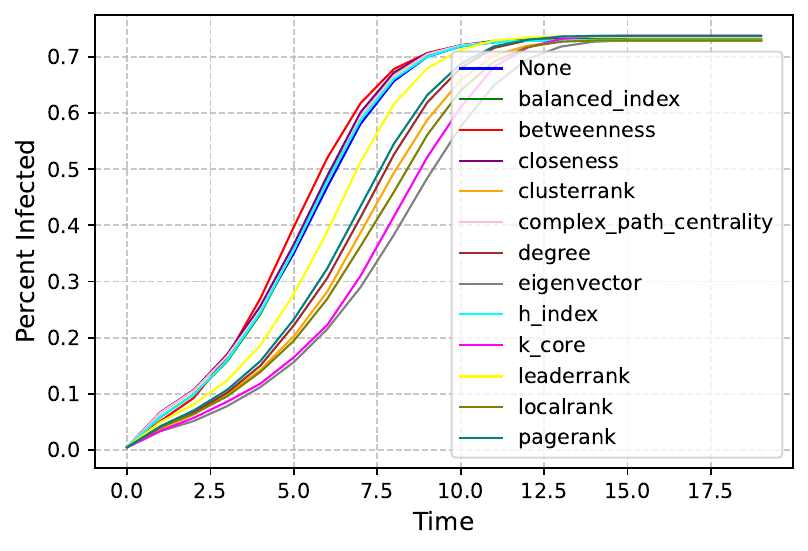}
        \caption{RA-2}
    \end{subfigure}
    \begin{subfigure}[b]{0.40\textwidth}
        \includegraphics[width=\textwidth]{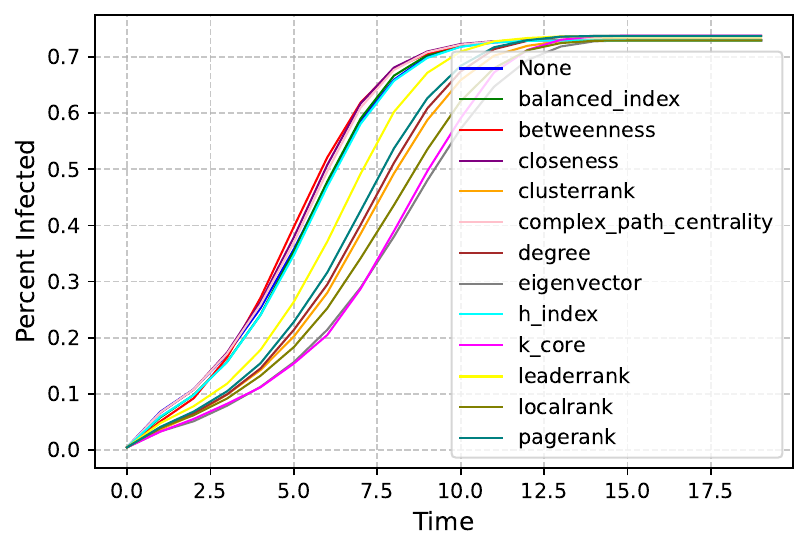}
        \caption{Jaccard}
    \end{subfigure}
    \caption{Centrality-VoteRank performance over predicted graphs for complex contagion (Part 1).}
    \label{fig:70_25_cagrqc_complex_part1}
\end{figure*}

\begin{itemize}
    \item \textbf{RA-2 consistently improves path-based centralities.} In Fig.~\ref{fig:70_25_cagrqc_complex_part1}(c), betweenness reaches 50\% infection one step earlier than in the original graph and outperforms all others by the end. This confirms that RA-2 predicts edges that enable better distance-based seed placement.

    \item \textbf{Some predicted graphs delay infection but converge similarly.} Jaccard’s sparser completions (Fig.~\ref{fig:70_25_cagrqc_complex_part1}d) yield curves that lag slightly but ultimately match the baseline. This suggests a tradeoff between cautious edge completion and seeding efficiency.

    \item \textbf{Local-structure-favoring metrics boost early spread.} In Fig.~\ref{fig:70_25_cagrqc_complex_part2}, Local Path and Quasi-Local RA improve early performance for closeness and betweenness. Under Resource Allocation (Fig.~\ref{fig:70_25_cagrqc_complex_part2}d), LeaderRank and closeness show earlier jumps, suggesting these completions densify the neighborhoods needed to initiate cascades.

    \item \textbf{Latent influencers emerge from completions.} In many predicted graphs, new nodes—previously invisible in the partial graph—are selected as seeds and contribute substantially to spread. This is clearest in RA-2 and Quasi-Local completions, where paths are extended in meaningful ways.
\end{itemize}

Finally, when examining full rankings across all centrality metrics and prediction strategies (see Appendix~\ref{sec:completetables}), several key trends appear. Metrics such as betweenness, closeness, complex path, and LocalRank routinely perform among the best, likely due to their sensitivity to network distances—especially critical under our deterministic, threshold-based contagion models.

In summary, applying Social Sphere predictions enables centrality-based algorithms to operate more effectively, revealing influential nodes that would otherwise remain hidden in partial graphs. This dual benefit—preserving known influencers while surfacing latent ones—is central to the power of our model.

\section{Conclusion} \label{sec:conclusion}

In this study, we extended and evaluated the \textbf{Social Sphere Model}, a framework that combines link prediction and influencer identification to forecast influential nodes in incomplete or evolving networks. Building on our prior work in~\cite{lin2025socialspheremodelheuristic}, we applied the Social Sphere framework to a real-world scientific collaboration network (arXiv GR-QC) and assessed its performance across a suite of contagion models and prediction strategies. The RA-2 similarity metric in particular showed consistently low mean squared error and high accuracy in influencer prediction, highlighting the model’s strength in uncovering latent influence potential. This work serves as a direct empirical continuation of our earlier theoretical and simulation-based study, further validating the model’s predictive capabilities in real-world settings.

By modifying classical influencer algorithms—most notably VoteRank—to operate over centrality rankings, we demonstrated that centrality-based variants such as those using betweenness or closeness often match or exceed the original performance. Our analysis further revealed that the choice of link prediction metric plays a crucial role: conservative metrics like Common Neighbors or Jaccard preserve known structure and are more reliable for short prediction horizons, while aggressive metrics like RA-2 or Quasi-Local RA add valuable reach when the snapshot is sparse or the time window $t$ is larger. These findings illustrate how different graph completion strategies uncover different types of influencers.

Beyond academic settings, the Social Sphere Model has applications in marketing, information dissemination, epidemiology, and influence forecasting. For instance, in viral marketing, applying our approach to partially observed social media networks can help anticipate future key influencers early enough to support negotiation, recruitment, or targeted interventions. The low computational cost of many centrality-based heuristics also makes the framework practical for large-scale use. Additionally, our findings suggest promise for improved network approximation and the detection of hidden influencers who may not yet be visible in incomplete or current data.

Finally, our current use of deterministic contagion processes leaves room to explore more realistic dynamics in future work. Extensions could include complex contagion models with reinforcement thresholds or probabilistic cascades. Likewise, expanding the pool of link prediction methods—such as incorporating clustering coefficients~\cite{clusteringcoeff} or applying iterative predictions as in~\cite{futurecommonneighbors}—could enhance flexibility and accuracy. Algorithm~\ref{alg:futureinfluencers} allows for tunable forecasting windows via $t$, but future edges induced by predicted links remain unmodeled. A more selective inclusion of edges or iterative updates may address this limitation and benefit structurally sensitive algorithms like LIR or Graph Coloring. Future studies could also evaluate Social Sphere across other network types, including temporal, directed, and weighted graphs, to identify combinations of predictors and influencers that are most effective under varying conditions, and compare with recent studies of collaboration networks (e.g. see \cite{new11, new22}).

\bigskip

\noindent {\bf Acknowledgments.} 
We extend our gratitude to the MIT PRIMES-USA program for their invaluable support and guidance throughout this research. The research of LPS was partially supported by  NSF FRG Award DMS- 2152107, a Simons Fellowship, an NSF CAREER Award DMS 1749013 and a Visiting Fellowship at All Souls College, Oxford.  \\

\noindent {\bf Affiliations.}
\begin{enumerate}
    \item[(a)] Harvard University, USA.
    \item[(b)] University of Illinois at Chicago,  USA.
    \item[(c)] Massachusetts Institute of Technology, USA.
\end{enumerate}
\bibliographystyle{plain}
\bibliography{bibliography}

\appendix
\section{More Data}\label{sec:completetables}
This appendix provides additional visualizations and data supporting the results presented in the main text. The figures below illustrate algorithm performance across various link prediction strategies and highlight the consistency of the Social Sphere Model across settings and metrics.

\begin{figure*}[t]
    \centering
    \begin{subfigure}[b]{0.45\textwidth}
        \centering
        \includegraphics[width=\textwidth]{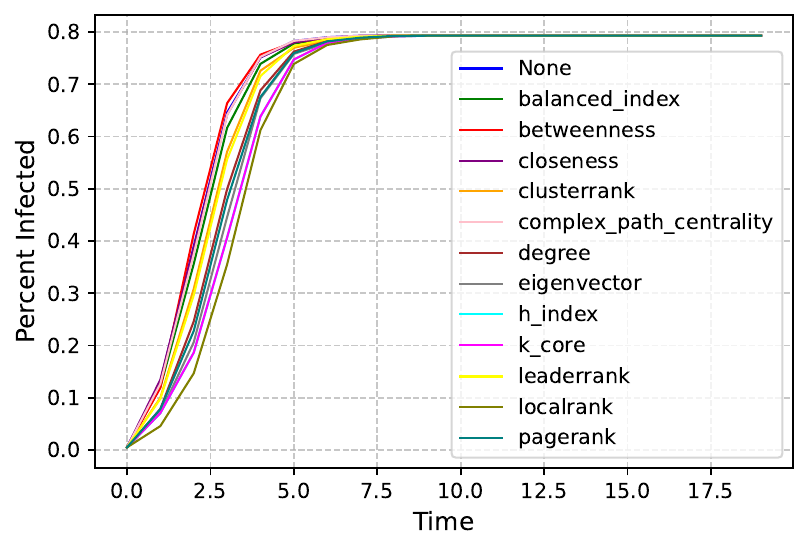}
        \caption{Original Graph}
    \end{subfigure}
    \begin{subfigure}[b]{0.45\textwidth}
        \centering
        \includegraphics[width=\textwidth]{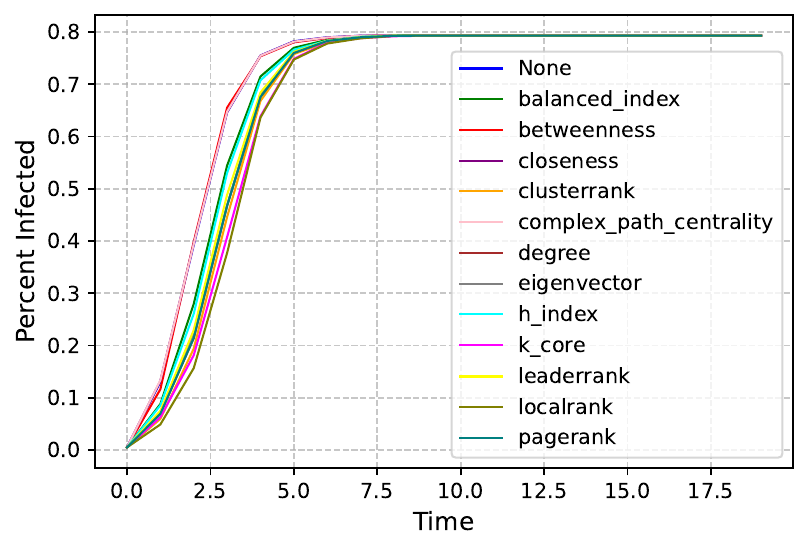}
        \caption{Common Neighbors Predicted Graph}
    \end{subfigure}
    \begin{subfigure}[b]{0.45\textwidth}
        \centering
        \includegraphics[width=\textwidth]{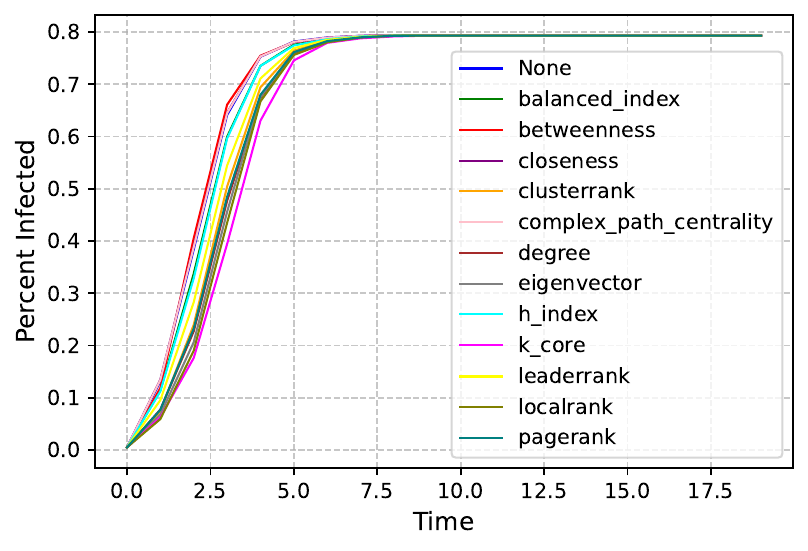}
        \caption{Jaccard Predicted Graph}
    \end{subfigure}
    \begin{subfigure}[b]{0.45\textwidth}
        \centering
        \includegraphics[width=\textwidth]{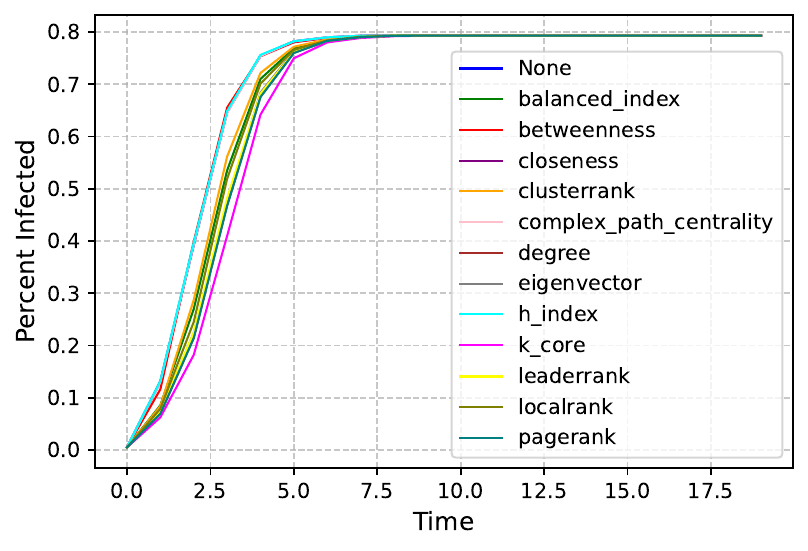}
        \caption{Local Path Predicted Graph}
    \end{subfigure}
    \begin{subfigure}[b]{0.45\textwidth}
        \centering
        \includegraphics[width=\textwidth]{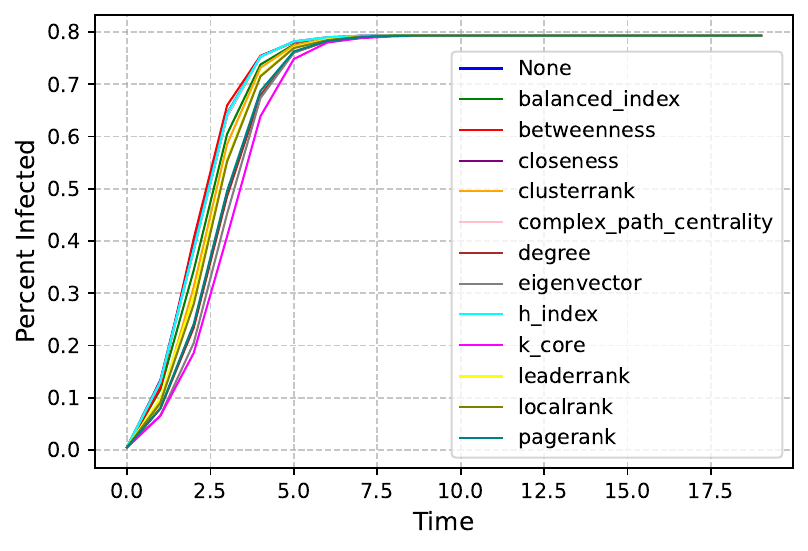}
        \caption{Quasi-Local RA Predicted Graph}
    \end{subfigure}
    \begin{subfigure}[b]{0.45\textwidth}
        \centering
        \includegraphics[width=\textwidth]{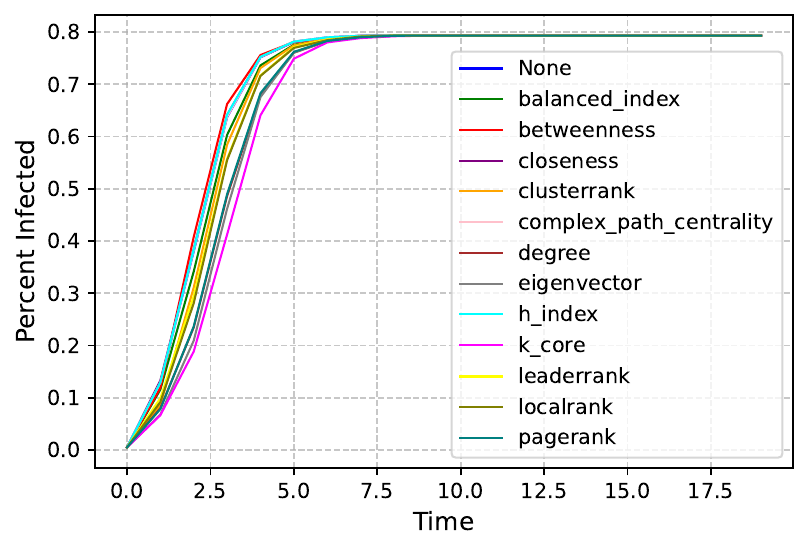}
        \caption{Quasi-Local RA-2 Predicted Graph}
    \end{subfigure}
    \begin{subfigure}[b]{0.45\textwidth}
        \centering
        \includegraphics[width=\textwidth]{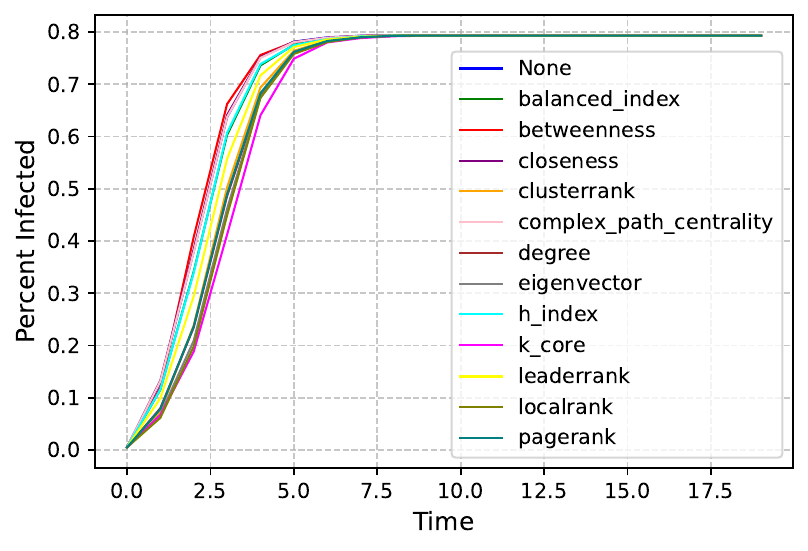}
        \caption{RA-2 Predicted Graph}
    \end{subfigure}
    \begin{subfigure}[b]{0.45\textwidth}
        \centering
        \includegraphics[width=\textwidth]{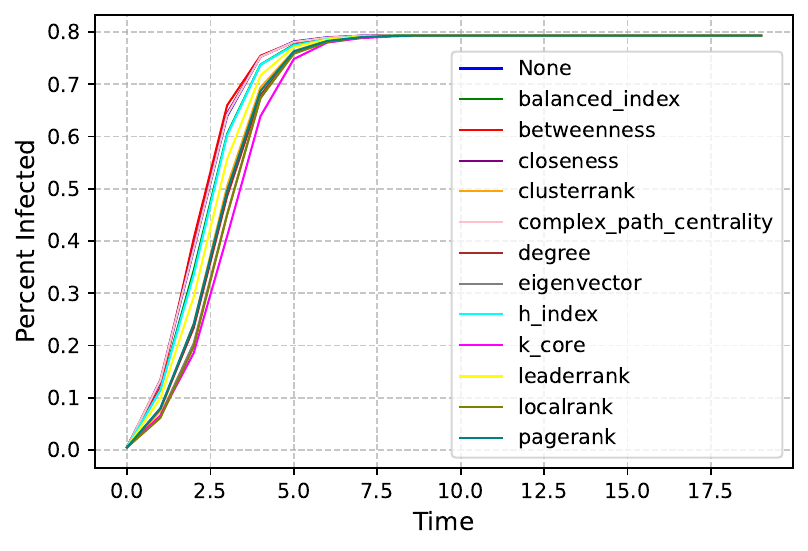}
        \caption{Resource Allocation Predicted Graph}
    \end{subfigure}
    \caption{Performance of influencers chosen with Centrality VoteRank (the original VoteRank is labelled as None) on predicted graphs formed from $70\%$ training graphs from the arXiv GrQc dataset in simple contagion.}
    \label{fig:70_25_cagrqc_simple}
\end{figure*} 

\begin{figure*}
    \centering
    \includegraphics[width=1\textwidth]{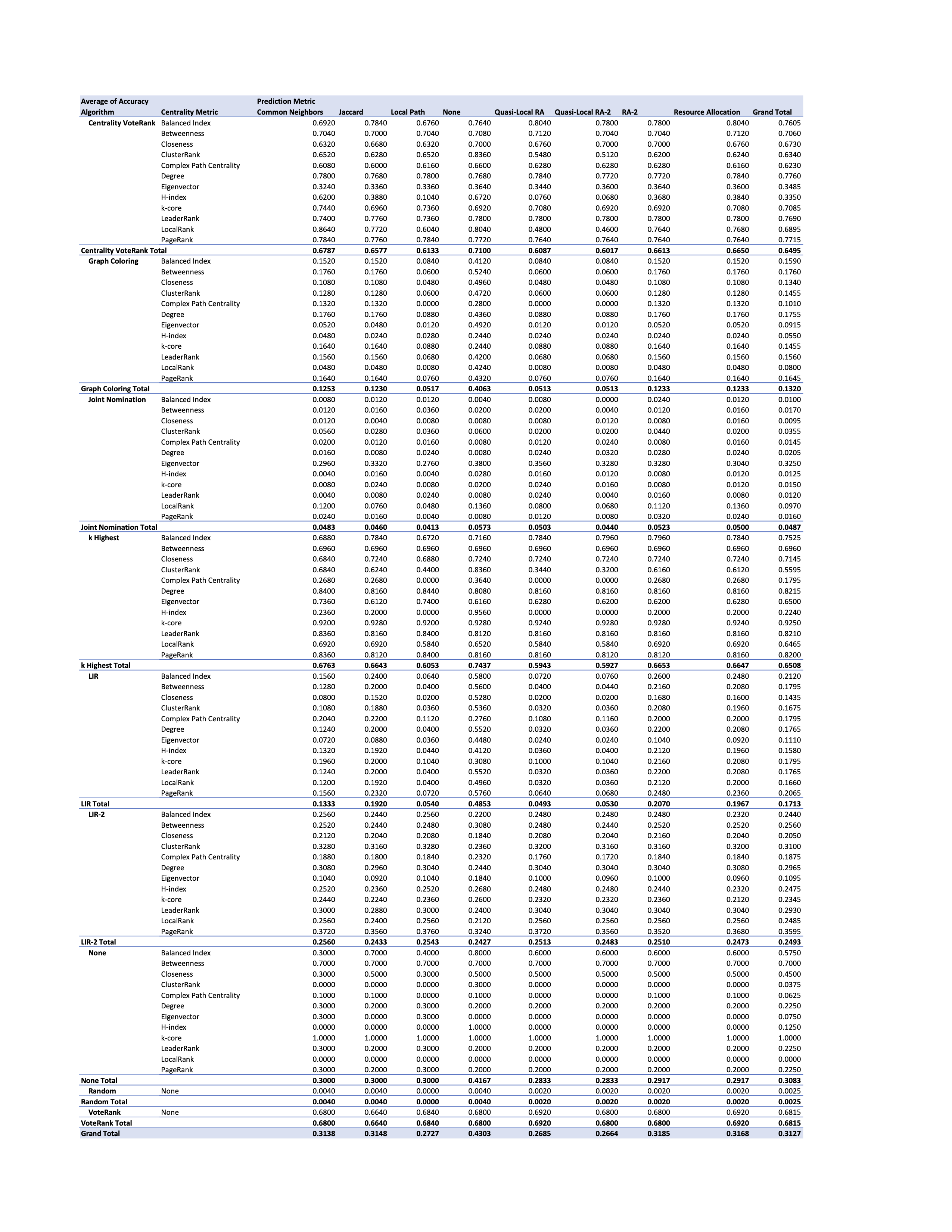}
    \caption{Average accuracies for different algorithms on $70\%$ training graphs from the arXiv GrQc dataset.}
    \label{fig:cagrqc_70_sims}
\end{figure*}
\begin{figure*}
    \centering
    \includegraphics[width=1\textwidth]{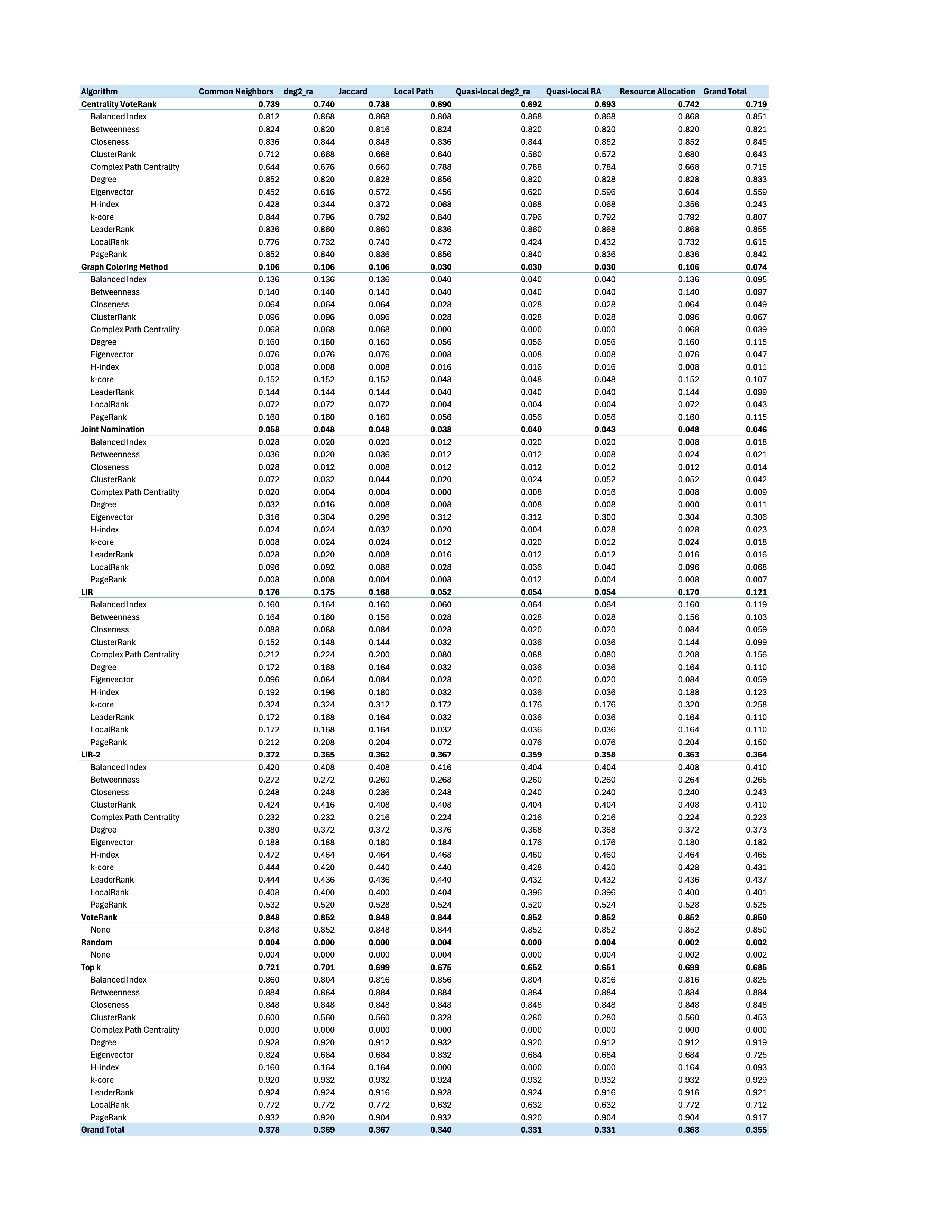}
    \label{fig:cagrqc_90_sims}
    \caption{Average accuracies for different algorithms on $90\%$ training graphs from the arXiv GrQc dataset.}
\end{figure*}

\end{document}